\documentclass[aps,prl,twocolumn,superscriptaddress,footinbib,nopacs,letterpaper]{revtex4-1}

\usepackage[colorlinks=true,linkcolor=blue,urlcolor=blue,citecolor=blue,anchorcolor=blue]{hyperref}
\usepackage{amsmath}
\usepackage{graphicx}
\usepackage{bm}
\usepackage{color}
\usepackage{mathbbol}
\usepackage{color}
\newcommand{\Ord}{\mathrm{O}}

\usepackage{pdfpages} % include pdfs
\usepackage{pgffor} % for loops

\makeatletter
\AtBeginDocument{\let\LS@rot\@undefined}
\makeatother

\def\supplementfilename{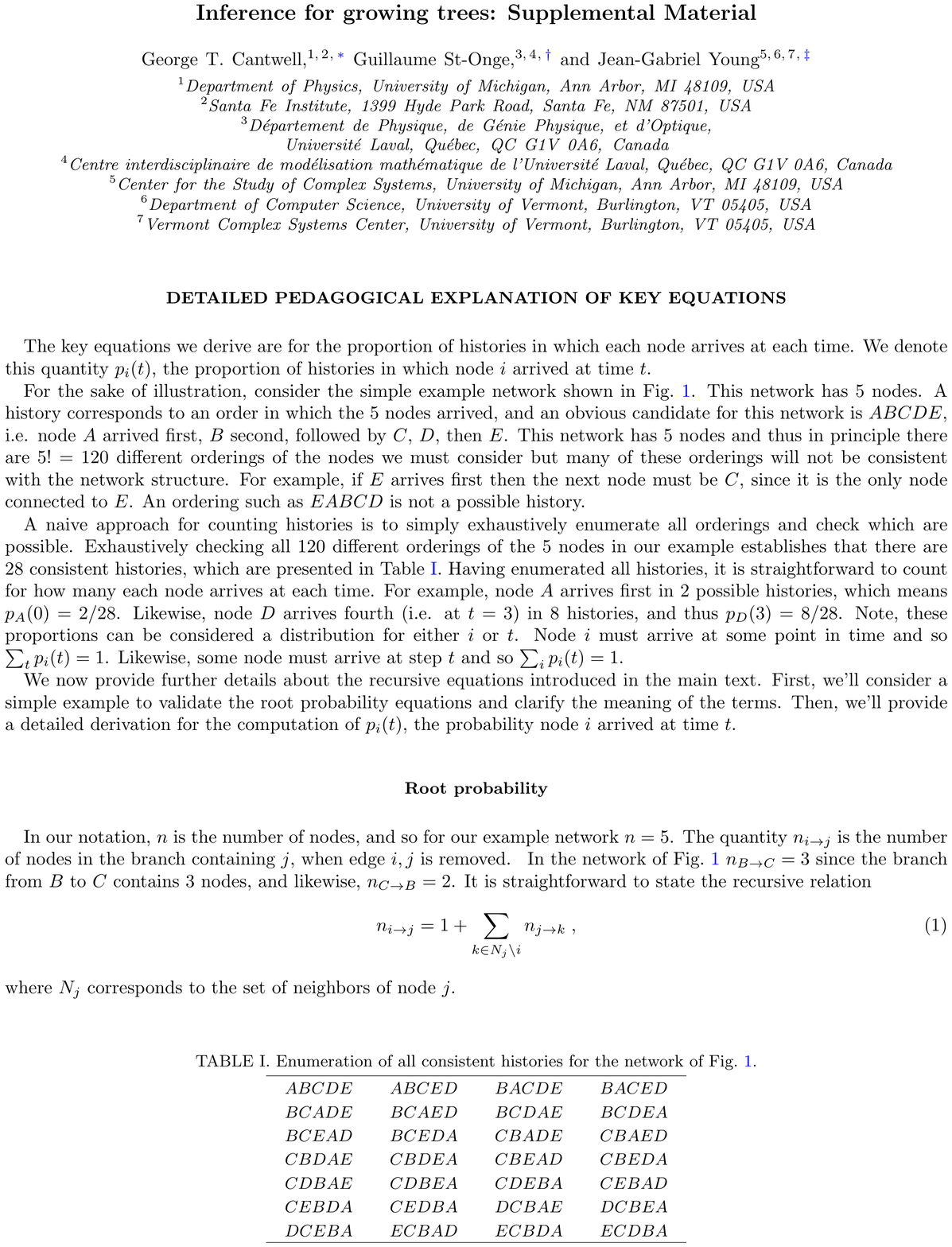}

\pdfximage{\supplementfilename}
\def\numbersupplementpages{\the\pdflastximagepages}

\newif\ifarXiv
\arXivtrue 
\begin{document}

\title{Inference for growing trees}

\author{George T. Cantwell}
\email[]{gcant@umich.edu}
\affiliation{Department of Physics, University of Michigan, Ann Arbor, MI 48109, USA}
\affiliation{Santa Fe Institute, 1399 Hyde Park Road, Santa Fe, NM 87501, USA}

\author{Guillaume St-Onge}
\email[]{guillaume.st-onge.4@ulaval.ca}
\affiliation{D\'epartement de Physique, de G\'enie Physique, et d'Optique, Universit\'e Laval, Qu\'ebec, QC G1V 0A6, Canada}
\affiliation{Centre interdisciplinaire de mod\'elisation math\'ematique de l'Universit\'e Laval, Qu\'ebec, QC G1V 0A6, Canada}

\author{Jean-Gabriel Young}
\email{jean-gabriel.young@uvm.edu}
\affiliation{Center for the Study of Complex Systems, University of Michigan, Ann Arbor, MI 48109, USA}
\affiliation{Department of Computer Science, University of Vermont, Burlington, VT 05405, USA}
\affiliation{Vermont Complex Systems Center, University of Vermont, Burlington, VT 05405, USA}

\begin{abstract}
One can often make inferences about a growing network from its current state alone. For example, it is generally possible to determine how a network changed over time or pick among plausible mechanisms explaining its growth. In practice, however, the extent to which such problems can be solved is limited by existing techniques, which are often inexact, inefficient, or both. In this Letter, we derive exact and efficient inference methods for growing trees and demonstrate them in a series of applications: network interpolation, history reconstruction, model fitting, and model selection.
\end{abstract}

\maketitle

A detailed description of the history of a complex network is highly informative.
For example, epidemic forecasting \cite{masuda2017introduction}, structural inference \cite{ghasemian2016detectability} or growth modeling \cite{overgoor2019choosing} all benefit from such descriptions.
Unfortunately, complete temporal descriptions are often unavailable \cite{reeves2019network}. 
The information may have been lost entirely---for instance the networks of ancient civilizations \cite{brughmans2010connecting}, or biological networks that have undergone evolution \cite{navlakha2011network,pinney2007reconstruction}.
Or, the complete temporal description may be too difficult to obtain---for instance ecosystems that cannot be sampled in real time \cite{kaiser2017ecosystem}, or very large networks whose evolution is too costly to track \cite{backstrom2006group}.

One may worry whether rigorous inference about the evolution of these networks is even possible when the temporal data is missing.
Recent research shows that is is possible, at least when a network is ``grown''---a process whereby nodes and edges are added to the network but never removed \cite{shah2011rumors,magner2017recovery,bubeck2017finding,lugosi2018finding,sreedharan2019inferring,wiuf2006likelihood,guetz2011adaptive,bloem2016random,Young2018}.
So long as one can reconstruct the temporal evolution of a network, inferences can be made.

Existing temporal reconstruction methods fall into three categories.
First, and simplest, are the methods that exploit correlations between the age of nodes and their properties (degree, centrality, etc.) \cite{magner2017recovery,Young2018}.
Second, and more sophisticated, are methods that use combinatorial techniques to identify the initial network or seed, a form of partial temporal reconstruction \cite{bubeck2017finding,lugosi2018finding}.
And third are Monte Carlo methods that rely on indirect sampling to reconstruct the complete history of statically observed networks \cite{wiuf2006likelihood,guetz2011adaptive,bloem2016random,Young2018,sreedharan2019inferring}.
All of these approaches achieve some form of temporal reconstruction
but they are also all imperfect: inferences based on correlations are imprecise \cite{Young2018}, the combinatorial methods do not yield complete reconstructions, and indirect sampling is not scalable \cite{bloem2016random,Young2018}.
Our analyses address these problems simultaneously.

We consider the case of growing trees: undirected networks that are generated by a process in which nodes arrive sequentially and attach to the extant network with a single edge.
When we have no information about the history of such trees (i.e., the order in which nodes arrived) temporal reconstruction amounts to picking from factorially many possible orderings of the nodes, many of which are, and many of which are not, consistent with the final structure of the network.

We introduce two distinct but related methods to make inferences about growing trees when temporal data is missing.
The first is a counting argument that provides the number of possible histories in which each node arrived at each time.
Such counts can be used to compute the most likely arrival time for each node, to establish our uncertainty about this quantity, and so forth.
The second method builds on the first by providing a Monte Carlo algorithm for directly sampling from the set of possible histories.
This algorithm allows any quantity of interest to be efficiently averaged over the set of all possible histories.
Of particular note, it allows us to evaluate the likelihood of different network growth models.

Our two methods (counting and direct sampling) assign an equal weight to all possible histories, and in this sense provide model-agnostic or nonparametric inferences.
Some well-known growth models, such as preferential attachment \cite{barabasi1999emergence}, induce a uniform posterior distribution over all consistent histories \cite{drmota2009random, Young2018},
and so counting arguments provide exact inferences for these models.
For more complicated models with non-uniform distributions over histories, one can use our Monte Carlo method together with standard reweighting techniques.

So, how do we begin accounting for the possible histories of a growing tree?
Let us start with the counting argument.
Our first goal will be to compute $p_i$, the proportion of histories consistent with the final network in which node~$i$ appears first.
As it turns out, there already exists an algorithm that computes $p_i$, introduced in Ref.~\cite{shah2011rumors} for a different purpose.
After describing the algorithm below,  we will prove that it correctly computes $p_i$ for any grown tree.

The algorithm proceeds as follows.
Suppose we have an undirected tree, $G$, with $n$ nodes.
First, arbitrarily root $G$ at any node.
For simplicity, pick node~$0$ to obtain the rooted tree $G_0$---a directed tree with the same edges as $G$ but with each edge directed to point away from node~$0$.
Next, for each directed edge $i \to j$ in $G_0$ we compute $n_{i \to j}$, the total number of descendants of node~$j$, including itself.
This can be written 
\begin{equation}
	n_{i \to j} = 1 + \sum_{k \in N_j \setminus i} n_{j \to k}
	\label{eq:n_i_j}
\end{equation}
where the sum is over descendants of node~$j$, or equivalently all neighbors of $j$ except $i$, denoted $N_j \setminus i$.
These numbers can be computed for all edges in linear time using recursion.

To obtain $p_i$ from the numbers $n_{i \to j}$ first set $p_0=1$. Then, starting at node $i=0$ set
\begin{equation}
	p_j = p_i \left( \frac{n_{i \to j}}{n - n_{i \to j}} \right)
	\label{eq:p_j_update}
\end{equation}
for each child~$j$ of node~$i$, and so forth down the directed tree, $G_0$.
A recursion again completes this in linear time.
Once this has finished we normalize $p$ so that $\sum_i p_i = 1$, which completes the calculation.

\begin{figure}
\centering
\includegraphics[width=0.9\columnwidth]{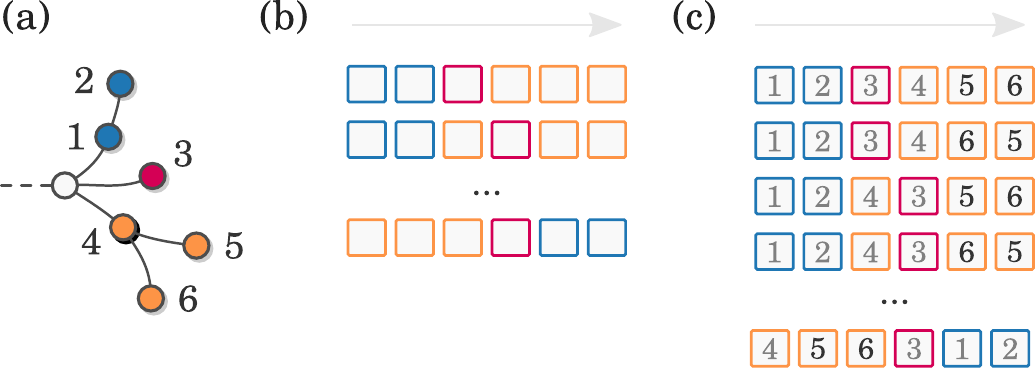}
\caption{
	Interlacing of histories.
	(a) We count the number of histories for the tree rooted on the focal node, shown in white.
	(b) There are $\frac{6!}{1! 2! 3!}=60$  ways to interlace the nodes of the 3 sub-trees, and (c) $1\times 1\times 2=2$ distinct orderings of these nodes for each of the interlacings.
    The total number of possible histories for the sub-tree rooted on the white node is therefore equal to $2\times 60=120$.
    See main text for the general combinatorial argument.
}
\label{fig:explanation}
\end{figure}

To see that Eq.~\eqref{eq:p_j_update} provides the correct proportions, one should imagine exhaustively enumerating all possible histories.
Let $h_i$ denote the total number of histories in which node~$i$ arrives first (thus, $p_i= h_i / \sum_j h_j$).
To calculate $h_i$, we imagine rooting the tree at node~$i$.
If we now removed node~$i$, the network would split into multiple sub-trees, each rooted at one of~$i$'s children.
Suppose, for the sake of the argument, that for each child node~$j$ we already knew $h_{i \to j}$, the number of histories in which $j$ is the seed of its sub-tree.
A  combinatorial argument then provides us with $h_i$.

To compute $h_i$ from $h_{i \to j}$, note that the nodes in each sub-tree can arrive in any of $h_{i \to j}$ orders, for a total of $\prod_{j} h_{i \to j}$ sub-histories across all branches.
However, this does not directly give us the number $h_i$, since multiple histories can be constructed by ``interlacing'' the same set of sub-histories in different ways (see Fig.~\ref{fig:explanation}).
The number of distinct interlacings is a multinomial coefficient and thus the number we actually want is 
\begin{equation}
	h_i =  (n - 1)! \prod_{j \in  N_i}  \frac{ h_{i \rightarrow j}}{n_{i \rightarrow j}!}.
	\label{eq:h_i}
\end{equation}

The quantity $h_{i \to j}$ can be computed in the same manner, but in the sub-tree with node $j$ as its root,
\begin{equation}
	h_{i \to j} =  (n_{i \to j} - 1)! \prod_{ k \in N_j \setminus i}  \frac{ h_{j \to k}}{n_{j \to k}!} ,
	\label{eq:h_i_j}
\end{equation}
where the product is now over children of $j$.
Equation~\eqref{eq:h_i_j} defines a set of self-consistent equations that can be solved directly using recursion and then substituted into Eq.~\eqref{eq:h_i}.
However, to find $p_i$ this is actually unnecessary.
If nodes~$i$ and $j$ are neighbors then from Eqs.~\eqref{eq:h_i} and \eqref{eq:h_i_j} 
\begin{equation}
	h_j = h_{i \to j} h_{j \to i}  \ \frac{(n-1)!}{(n_{i \to j} -1)!\ n_{j \to i}!},
\end{equation}
and from the equivalent expression for $h_i$ we see that for neighbors~$i$ and $j$,
\begin{equation}
	\frac{p_j}{p_i} = \frac{h_j}{h_i} = \frac{n_{i \to j}}{n_{j \to i}} = \frac{n_{i \to j}}{ n - n_{i \to j}}.
\end{equation}
This completes the proof that Eq.~\eqref{eq:p_j_update} is indeed correct.

This calculation is an important first step but we are far from finished---we can currently account for only one step of the history.
So, we now derive expressions for $p_i(t)$, the proportion of histories in which node~$i$ arrives at time~$t$, making use of the quantities $n_{i \to j}$ and $h_{i \to j}$ already defined in Eqs.~\eqref{eq:n_i_j} and~\eqref{eq:h_i_j}.
The combinatorial arguments we develop to derive these quantities are the key to inference on growing trees \footnote{See Supplemental Material at [URL will be inserted by publisher] for  a pedagogical derivation of the method and additional details.}.

If node~$i$ arrived at time~$t>0$, then precisely one of its immediate neighbors must have arrived before $t$.
We can use this fact to compute $p_i(t)$ by first counting the number of histories in which node $i$ arrives at time $t$ \emph{and} node $j \in N_i$ arrives before $t$; summing the resulting quantities for all $j\in N_i$ provides $p_i(t)$.

So, suppose for the sake of argument that it was node~$j$ that arrived before node~$i$.
Edge~$(i,j)$ separates two branches of the network and we consider the histories for each branch separately.
In the branch containing node~$i$, $i$ must arrive first, and there are $h_{j \to i}$ histories in which this occurs.
For the branch containing node~$j$, let $g_{i \to j}(t)$ denote the number of histories for which node~$j$ arrives before $t$.
To combine the histories in each branch (and thus count full histories of the original tree), note that the first $t$ nodes to arrive must be in $j$'s branch, and the next node is $i$ itself.
The number of ways of interlacing the remaining $n_{i \to j} - t$ nodes in $j$'s branch with the $n_{j \to i} - 1$ in $i$'s branch is a binomial coefficient.
Thus, the total number of histories in which $j$ arrives before $t$ and $i$ arrives at exactly $t$ is $g_{i \to j}(t) h_{j \to i} {n-t-1 \choose n_{i \to j} - t}$.
Summing this quantity over all of $i$'s neighbors
\begin{equation}
	p_i(t) = \frac{1}{Z} \sum_{j \in N_i} g_{i \to j}(t) h_{j \to i} {n-t-1 \choose n_{i \to j} - t},
	\label{eq:p_it}
\end{equation}
where $Z$ is the total number of histories consistent with $G$, which ensures $\sum_i p_i(t) = \sum_t p_i(t) = 1$.

Our calculation is currently incomplete since we do not yet know $g_{i \to j}(t)$, i.e., the number of histories in $j$'s branch in which it arrives before $t$.
However, the number of histories in which $j$ arrives at exactly time $t$ is $g_{i \to j}(t+1) - g_{i \to j}(t)$, and this can be calculated using the same argument as before.

For node~$j$ to arrive at time~$t>0$ in $j$'s branch (when edge~$(i,j)$ is removed), precisely one of $j$'s neighbors in this branch must have arrived before $t$.
Suppose for now that it was node~$k$.
We again consider the branches containing $j$ and $k$ separately.
There are $g_{j \to k}(t)$ histories where node $k$ arrives before time $t$ in $k$'s branch.
Each of these can be interlaced with any one of the $h_{i,k \to j}$ histories starting at $j$ in its branch once both edge $(i,j)$ and $(j,k)$ are removed.
Hence
\begin{multline}
	g_{i \to j}(t+1) - g_{i \to j}(t)\\ = \sum_{k \in N_j \setminus i} g_{j \to k}(t) h_{i,k \to j} {n_{i \to j} -t -1 \choose n_{j \to k} - t},
	\label{eq:g_i_diff}
\end{multline}
where the sum is over the neighbors of  $j$ except $i$, and the quantity $h_{i,k \to j}$ is calculated similarly to Eq.~\eqref{eq:h_i_j}, as
\begin{align}
	h_{i,k \to j} &= (n_{k \to j} - 1 - n_{j \to i})! \prod_{l \in N_j \setminus i,k} \frac{h_{j \to l}}{n_{j \to l}!} \nonumber \\
	&= \frac{ h_{k \to j} n_{j \to i}! (n_{k\to j}-1-n_{j\to i})! }{h_{j \to i} (n_{k\to j}-1)!}.
	\label{eq:last_equation}
\end{align}
These expressions only depend on $n_{i \to j}$ and $h_{i \to j}$, which we already know.
The result is that we now have a complete set of self-consistent equations for $p_i(t)$, for all $i$ and $t$.
We solve these equations as follows \footnote{ 
Although our notation is reminiscent of belief propagation \cite{Mezard2009}, it should be noted that the equations in fact differ.  Since this problem does not ``factorize'', the belief propagation formalism cannot be applied exactly nor efficiently.}.
First, set $g_{i \to j}(1) = h_{i \to j}$.
Then, for each $t>1$, use Eq.~\eqref{eq:g_i_diff} to compute $g_{i\to j}(t+1)$ from $g_{i\to j}(t)$.
Finally, set $p_i(0)=p_i$ and for each $t > 0$ set $p_i(t)$ using Eq.~\eqref{eq:p_it}.
The normalizing factor $Z$ is computed by requiring $\sum_i p_i(t)=1$.
Each iteration of Eq.~\eqref{eq:g_i_diff} takes $\Ord(q n)$ time where \mbox{$q=(\langle k^2 \rangle - \langle k \rangle) / \langle k \rangle $}, the ``excess degree'', with $\langle k \rangle$ being the average degree.
The total run time is therefore $\Ord( q n^2 )$.

This calculation is exact and reasonably efficient. We are computing $n^2$ quantities in $q n^2$ operations. However, for many purposes all $n^2$ values of $p_i(t)$ are more than we need. 
For example, we might only need to know the expected arrival times, $\langle t_i \rangle = \sum_t t p_i(t)$, which are optimal estimators for true arrival times \cite{Young2018}.
To complement our analytic method we now introduce an efficient Monte Carlo procedure to sample directly from the set of histories.

Since we can already compute the seed probability $p_i$ in linear time, we need to solve an interpolation problem.
Namely, we need to find a method that can generate uniform samples from the set of histories that start at an initial network $G_{I}$ and end at a final network $G_{F}$, ``bridging'' between the two states \cite{bloem2016random}.
Sampling uniformly from histories is then achieved by picking the seed proportional to $p_i$ and interpolating between this node and $G$.

For the class of growth processes we are considering all histories progress by attaching one node to the extant network.
As a result, when the network has $t$ nodes, the next node to arrive must be directly attached to one of these $t$ nodes in the final network.
In other words, the next node must be chosen from the ``boundary'' set  $B_t$, the set of all nodes that are not yet in the network but in the final state are directly connected to one that is.

To generate a complete bridge between $G_I$ and $G_F$ we can, without loss of generality, root $G_F$ at any node in $G_I$ and compute $n_{j \to k}$ in the rooted tree.
All nodes (except the root) will have an in-degree of $1$ so for simplicity we write $n_k=n_{j\to k}$.
Then, at each time we sample a node from the boundary set proportional to $n_k$.
At time $t$ the probability of adding node $k \in B_t$ is
\begin{equation}
	\frac{n_k}{\sum_{l \in B_t} n_l }.
\end{equation}
Noting also that the denominator $\sum_{l \in B_t} n_l$ is equal to the number of nodes that are not yet in the network, which is $n-t$, we see that node~$k$ is added at time~$t$ with probability $n_k / (n-t)$.
A specific interpolation $H = v_{I+1},v_{I+2},\dots,v_{F}$ is generated with probability
\begin{equation}
	P(H) = \prod_{t={I+1}}^{F} \frac{n_{v_t}}{n-t},
\end{equation}
which is independent of $H$, and thus uniform over consistent histories.
By using an efficient set sampling algorithm we can both update and sample from $B_t$ in $\Ord( \log \log n)$ time  \cite{st2019efficient}. 
The full procedure is thus close to linear---$\Ord( n \log \log n)$.

Since our methods are efficient, we can use them to tackle temporal reconstruction tasks on large trees, opening up many applications. 

\begin{figure}[!t]
\centering
\includegraphics[width=0.8\columnwidth]{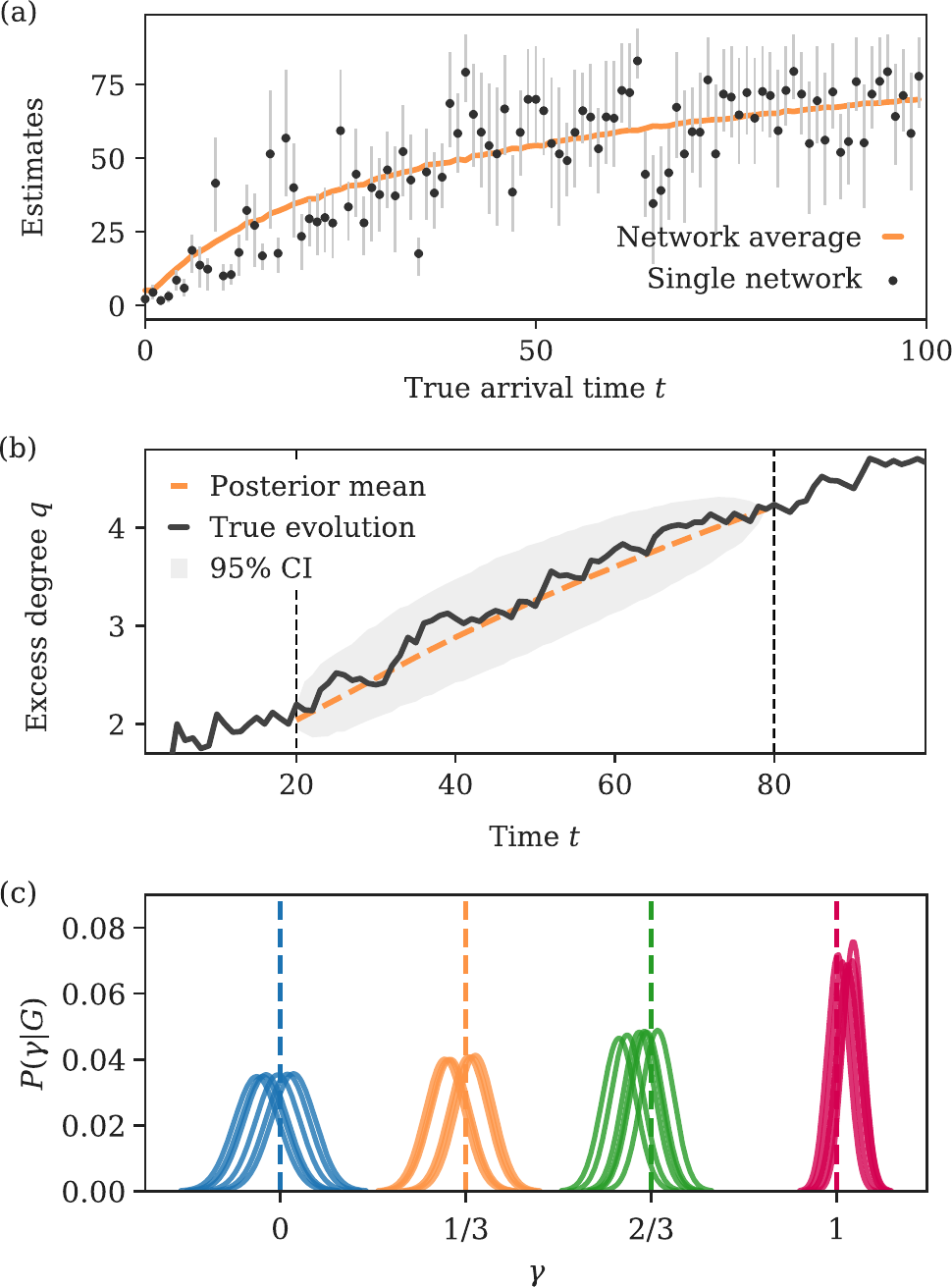}
\caption{ 
	Examples of inference problems solved by our methods.
	(a) Inference of the arrival time of nodes in networks of $100$ nodes grown with the uniform attachment rule, where incoming nodes attach to an extant node chosen uniformly at random.
 	The black symbols show the posterior mean of the arrival times $\langle \tau_i\rangle$ calculated for a single network, and the error bars denote the 50 \% credible interval interval (CI) for each $\tau_i$.
 	Both sets of quantities are calculated exactly with Eqs.~\eqref{eq:n_i_j}--\eqref{eq:last_equation}.
	Hence, randomness in the network growth process rather than the estimation procedure explain the fluctuations.
    They disappear once we average over networks, as shown by the solid line ($3\,000$ networks). As a comparison, these posterior mean estimators achieve an average correlation of $\approx 0.754$ with the true history, while estimators based on the degree centrality achieve $\approx 0.654$.
	(b) Evolution of the excess degree of a single preferential attachment network \cite{barabasi1999emergence} whose structure is known before $t=20$ and after $t=80$ but not in between.
	The shaded region shows the 95\% CI on an interpolation of the excess degree between these two observations, while the orange line shows the posterior mean.
	(c) Posterior distributions of $\gamma$ for networks generated with the generalized preferential attachment model and an attachment kernel $k^{\gamma}$ \cite{krapivsky2000connectivity}. 
	Each posterior distribution $P( \gamma \vert G)$ is computed from $100$ samples, and corresponds to a network of $n=1000$ nodes with kernel exponent  $\gamma = 0, 1/3, 2/3, 1$ (see main text).
}
\label{fig:fig2}
\end{figure}

In the network archaeology problem (Fig.~\ref{fig:fig2}a) the goal is to infer the past states of a network given only its final state \cite{navlakha2011network,Young2018}.
To make estimates of each node's arrival time, one can compute $\langle t_i \rangle = \sum_t t p_i(t)$ using Eq.~\eqref{eq:p_it}.
If we believe the network grew by some more complicated model (i.e., one with a non-uniform posterior distribution) we can use Monte Carlo samples to estimate
\begin{equation}
	\langle t_i \rangle = \sum_{H}  \tau_{i}(H) P(H),
\end{equation}
where $\tau_{i}(H)$ is the time of arrival of node $i$ in history $H$ and $P(H)$ is the probability of sequence $H$ under the chosen growth model.

In the network interpolation problem (Fig.~\ref{fig:fig2}b), the goal is to infer the evolution of the network between two known snapshots \cite{reeves2019network}.
One can again use the sampling method to interpolate between the initial state and final state, tracking the evolution of any network quantity of interest along the way.
Non-uniform models are handled by computing average trajectories with reweighting.

A second class of problems our methods solve is model inference.
For example, if we want to estimate a parameter $\gamma$, we must be able to compute the likelihood of the model
\begin{equation}
	P(G \vert \gamma) = \sum_{H} P(G, H \vert \gamma),
	\label{eq:likelihood_def}
\end{equation}
which corresponds to the probability of observing network $G$ given $\gamma$.
Our method allows us to compute the sum in Eq.~\eqref{eq:likelihood_def}, and by extension to make inferences about $\gamma$ (for example using the method of maximum likelihood, or Bayesian procedures).

Let's consider a concrete example: kernel inference \cite{Newman01d, JNB03, overgoor2019choosing, sheridan2012measuring}, in which $\gamma$ is a parameter that determines the structure of the tree.
We suppose that each time a new node is introduced, it connects to an extant node proportional to that node's degree raised to the power $\gamma$  \cite{krapivsky2000connectivity}.
To evaluate $\gamma$ from a network, we sample histories $\{H_1, H_2, \dots\}$ and then evaluate
\begin{equation} 
	\label{eq:likelihood_MC}
	P(\gamma \vert G) \propto \sum_{s} P(G,H_s \vert \gamma).
\end{equation}
Panel (c) of Fig.~\ref{fig:fig2} shows how we are able to recover the attachment kernel, given only the final network as input.

The same technique lets us test mechanisms, like preferential attachment, and determine whether they are plausible models for any given tree.
One might hope to carry out such test by inspecting simple network statistics, such as the degree distributions, but such methods fail when similar statistics are generated by different mechanisms.
For example, the degree distribution of the \emph{redirection} model \cite{krapivsky_organization_2001} can be remarkably similar to that of the preferential attachment \cite{Note1}.
But, while the degree distribution are similar, our methods can easily separate the models by accounting for the whole structure.
To test this we generated $100$ networks of $2\,048$ nodes from each model.
Bayes factor \cite{migon_statistical_2014}  correctly identified the right model in every single case with an average strength of evidence of $\langle \vert \log K\vert \rangle = 25.25$.

\begin{table}
\caption{95\% credible interval for $\gamma$, the exponent of the a non-linear preferential attachment kernel \cite{krapivsky2000connectivity} fitted to growing network data. The first  results in the middle column are computed with temporal information and the ones in the right column without.}
\begin{ruledtabular}
\begin{tabular}{lccc}
	Network  & $n$ & Known timeline & Reconstructed\\\hline
	Phylogenetic tree      & 4120 & $[ -0.41, -0.20]$& $[-0.52,-0.39]$\\
	Twitter reply tree      & 748 & $[0.93, 1.03]$& $[0.89, 1.00]$\\
	Erd\H{o}s collaborators & 6927 & ---& $[1.18, 1.21]$\\
\end{tabular}
\end{ruledtabular}
\label{table:CI} 
\end{table}

We end with a simple demonstration using empirical data (see Table \ref{table:CI}).
When applied to real trees without temporal meta-data our method finds that the co-authorship network centered on Paul Erd\H{o}s \cite{batagelj2009pajek} is plausibly grown by a super-preferential attachment mechanism \cite{krapivsky2000connectivity}, a network of re-tweets on Twitter \cite{garland_countering_2020} is explained by a regular preferential attachment mechanism, and the phylogenetic tree of Western Nile Virus \cite{hadfield2018nextstrain} certainly did not grow by this mechanism.
The credible intervals found when using only the static network overlap with the one we find when using available temporal meta-data. 

To conclude, we have developed analytic expressions and an efficient Monte Carlo method for inference problems for growing trees.
In artificial and real-world networks, we have shown that these methods can be used to make rigorous inference from statically observed networks.
This is an important step toward a statistically principled selection of growth mechanisms in network science, and their usage as generative models for temporal reconstruction.
One important challenge we leave open is inference for networks that are not trees.

\begin{acknowledgments}
The authors thank  Laurent H\'ebert-Dufresne, Alec Kirkley and Mark Newman for helpful discussions.
This work was funded in part by the Conseil de recherches en sciences naturelles et en g\'enie du Canada (GSO) and  the James S. McDonnell Foundation (JGY).
Code implementing our methods is available online at \url{github.com/gcant/temporal-recovery-tree-py}.
\end{acknowledgments}

\bibliography{references.bib}

\ifarXiv
    

\section*{Supplementary material}

# Inference for growing trees: Supplemental Material

George T. Cantwell,[1, 2, *] Guillaume St-Onge,[3, 4, †] and Jean-Gabriel Young[5, 6, 7, ‡]

[1]*Department of Physics, University of Michigan, Ann Arbor, MI 48109, USA*
[2]*Santa Fe Institute, 1399 Hyde Park Road, Santa Fe, NM 87501, USA*
[3]*Département de Physique, de Génie Physique, et d'Optique,
Université Laval, Québec, QC G1V 0A6, Canada*
[4]*Centre interdisciplinaire de modélisation mathématique de l'Université Laval, Québec, QC G1V 0A6, Canada*
[5]*Center for the Study of Complex Systems, University of Michigan, Ann Arbor, MI 48109, USA*
[6]*Department of Computer Science, University of Vermont, Burlington, VT 05405, USA*
[7]*Vermont Complex Systems Center, University of Vermont, Burlington, VT 05405, USA*

## DETAILED PEDAGOGICAL EXPLANATION OF KEY EQUATIONS

The key equations we derive are for the proportion of histories in which each node arrives at each time. We denote this quantity $p_i(t)$, the proportion of histories in which node $i$ arrived at time $t$.

For the sake of illustration, consider the simple example network shown in Fig. 1. This network has 5 nodes. A history corresponds to an order in which the 5 nodes arrived, and an obvious candidate for this network is $ABCDE$, i.e. node $A$ arrived first, $B$ second, followed by $C$, $D$, then $E$. This network has 5 nodes and thus in principle there are $5! = 120$ different orderings of the nodes we must consider but many of these orderings will not be consistent with the network structure. For example, if $E$ arrives first then the next node must be $C$, since it is the only node connected to $E$. An ordering such as $EABCD$ is not a possible history.

A naive approach for counting histories is to simply exhaustively enumerate all orderings and check which are possible. Exhaustively checking all 120 different orderings of the 5 nodes in our example establishes that there are 28 consistent histories, which are presented in Table I. Having enumerated all histories, it is straightforward to count for how many each node arrives at each time. For example, node $A$ arrives first in 2 possible histories, which means $p_A(0) = 2/28$. Likewise, node $D$ arrives fourth (i.e. at $t = 3$) in 8 histories, and thus $p_D(3) = 8/28$. Note, these proportions can be considered a distribution for either $i$ or $t$. Node $i$ must arrive at some point in time and so $\sum_t p_i(t) = 1$. Likewise, some node must arrive at step $t$ and so $\sum_i p_i(t) = 1$.

We now provide further details about the recursive equations introduced in the main text. First, we'll consider a simple example to validate the root probability equations and clarify the meaning of the terms. Then, we'll provide a detailed derivation for the computation of $p_i(t)$, the probability node $i$ arrived at time $t$.

### Root probability

In our notation, $n$ is the number of nodes, and so for our example network $n = 5$. The quantity $n_{i \to j}$ is the number of nodes in the branch containing $j$, when edge $i, j$ is removed. In the network of Fig. 1 $n_{B \to C} = 3$ since the branch from $B$ to $C$ contains 3 nodes, and likewise, $n_{C \to B} = 2$. It is straightforward to state the recursive relation

$$n_{i \to j} = 1 + \sum_{k \in N_j \setminus i} n_{j \to k} \, , \tag{1}$$

where $N_j$ corresponds to the set of neighbors of node $j$.

TABLE I. Enumeration of all consistent histories for the network of Fig. 1.

| | | | |
|---|---|---|---|
| $ABCDE$ | $ABCED$ | $BACDE$ | $BACED$ |
| $BCADE$ | $BCAED$ | $BCDAE$ | $BCDEA$ |
| $BCEAD$ | $BCEDA$ | $CBADE$ | $CBAED$ |
| $CBDAE$ | $CBDEA$ | $CBEAD$ | $CBEDA$ |
| $CDBAE$ | $CDBEA$ | $CDEBA$ | $CEBAD$ |
| $CEBDA$ | $CEDBA$ | $DCBAE$ | $DCBEA$ |
| $DCEBA$ | $ECBAD$ | $ECBDA$ | $ECDBA$ |

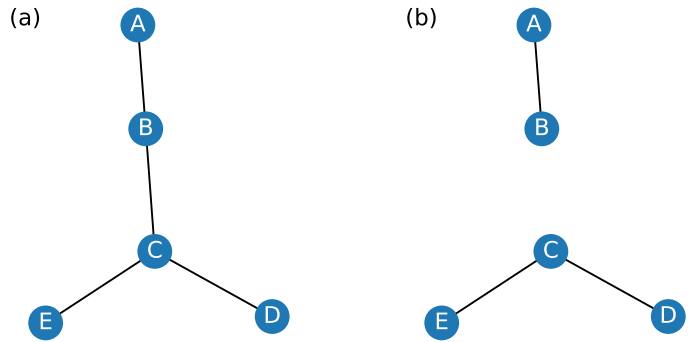

FIG. 1. In (a) we show an example of a network with 5 nodes. In (b) we show the two branches that are defined by removing edge $B, C$. The quantities $n_{C \to B} = 2$ and $n_{B \to C} = 3$ since there are 2 nodes in $B$'s branch and 3 nodes in $C$'s. Likewise, $h_{C \to B} = 1$ and $h_{B \to C} = 2$ since there is one history rooted at $B$ in its branch (namely $BA$), and two histories in $C$'s branch rooted at $C$ (namely $CDE$ and $CED$).

The quantity $h_{i \to j}$ is the number of histories in the branch containing node $j$, when edge $i, j$ is removed, and where node $j$ is the root. This branch has $n_{i \to j}$ nodes and so in principle there are $n_{i \to j}!$ different orderings of the nodes we must consider. But as before, not all of them are consistent with the network structure. If, for example, we remove edge $B, C$ in the network of Fig. 1, we see that the branch containing $B$ and $A$ has only one history in which $B$ is the root: $BA$. Whereas, the branch containing $C$ has two histories in which $C$ is the root: $CDE$ and $CED$. Thus, in our example $h_{C \to B} = 1$ and $h_{B \to C} = 2$. This is concordant with the general recursive expression

$$h_{i \to j} = (n_{i \to j} - 1)! \prod_{k \in N_j \setminus i} \frac{h_{j \to k}}{n_{j \to k}!}, \qquad (2)$$

discussed in detail in the main text. For instance, we would have

$$h_{C \to B} = (2-1)! \frac{h_{B \to A}}{1!} = 1, \qquad (3)$$

since $h_{B \to A}$ is necessarily 1.

In the main text we also provide an algorithm to compute the root probability $p_i \equiv p_i(0)$ without having to compute the $h_{i \to j}$ quantities, simply using the relation

$$\frac{p_j}{p_i} = \frac{n_{i \to j}}{n - n_{i \to j}} \qquad (4)$$

for adjacent nodes $i$ and $j$. Applying the algorithm to the network in Fig. 1, we would first arbitrarily set $p_A = 1$. Then, set $p_B = p_A \left( \frac{n_{A \to B}}{5 - n_{A \to B}} \right) = 4$, then $p_C = p_B \left( \frac{n_{B \to C}}{5 - n_{B \to C}} \right) = 6$, and $p_D = p_E = 3/2$. Finally, we normalize the proportions so that $p_A + p_B + p_C + p_D + p_E = 1$. The final answer is $p_A = 1/14$, $p_B = 4/14$, $p_C = 6/14$ and $p_D = p_E = 3/28$, which agrees with the enumeration in Table I.

### Probability for each node to arrive at each time

Conceptually, our equations for the proportions $p_i(t)$ follow the same logic as explicit enumeration—we calculate the number of histories in which each node arrived at each time, and then divide this by the total number of histories. In other words, we denote the number of histories in which node $i$ arrives at time $t$ as $h_i(t)$ and then calculate $p_i(t)$ as

$$p_i(t) = \frac{h_i(t)}{\sum_{t'} h_i(t')}. \qquad (5)$$

How do we calculate $h_i(t)$? First, we note that when node $i$ arrived at time $t > 0$, it must have connected to one of its neighbors in the final network. If we can compute the number of histories in which node $i$ arrived at time $t$ and connected to a neighbor node $j$, and then sum this quantity over all neighbors $j$, we will arrive at $h_i(t)$.

So, let's consider the scenario in which node $i$ arrived at time $t$ and connected to node $j$. As before, edge $i, j$ defines two branches of the tree—one branch containing node $i$ and the other, $j$. To compute the total number of histories in which node $i$ arrived at time $t$ and connected to node $j$, we consider the histories in each branch separately.

The branch containing node $i$ must have $i$ as its root, and the total number of histories (possible orderings of the $n_{j \to i}$ nodes) we must consider in this branch is by definition $h_{j \to i}$. The branch containing node $j$, however, need not have node $j$ as its root, and so $h_{i \to j}$ is not the relevant number. Nevertheless, node $j$ must have arrived before time $t$ (if it didn't, then node $i$ couldn't connect to $j$ at time $t$). Although we might not yet know this number, let $g_{i \to j}(t)$ denote the number of histories in the branch containing node $j$ in which node $j$ arrived before time $t$.

Returning to our example in Fig. 1, consider the scenario in which $B$ arrived at time 2 and connected to $C$. Edge $B, C$ defines two branches of the tree, one that contains $B$ and $A$, and another that contains $C$, $D$, and $E$. We first count the histories in these two branches separately.

Node $B$ must be the root of its branch which means there is only one possible history: $BA$. Node $C$, however, need not be the root of its branch. All that we require is for node $C$ to arrive before time 2. This leaves 4 possible histories: $CDE$, $CED$, $DCE$, and $ECD$, i.e. $g_{B \to C}(2) = 4$.

To produce a full history for the network, each of the $h_{j \to i}$ histories in the branch containing node $i$ can be combined which each of $g_{i \to j}(t)$ histories in the branch containing $j$. When combining any pair of histories from the two branches, the first $t$ steps must occur in the branch containing $j$, and the next step must be $i$ itself arriving. After this there will be $n_{j \to i} - 1$ nodes still to arrive in $i$'s branch and $n_{i \to j} - t$ nodes still to arrive in $j$'s branch. The $n_{i \to j} - t$ nodes from the history in $j$'s branch can be slot into any of the remaining $n - t - 1$ time steps, and the number of ways of doing this is the binomial coefficient, $\binom{n-t-1}{n_{i \to j}-t}$.

Returning yet again to our example, if $B$ connects to $C$ at time 2, then the first two steps must occur in the branch containing $C$. The four options for the first two steps are $CD$, $CE$, $DC$, or $EC$. Next, node $B$ itself arrives. This still leaves the last two steps undetermined. Each branch still has one node left to arrive, and we have two steps left to introduce them. There are total of $\binom{2}{1} = 2$ possibilities for the remaining nodes—we have to choose a time step for the remaining 1 node in $C$'s branch, and we have 2 steps available to choose from. Putting this all together, there are a total of

$$g_{B \to C}(2) h_{C \to B} \binom{2}{1} = 4 \times 1 \times 2 = 8 \tag{6}$$

histories in which $B$ arrives at time 2 and connects to $C$. Finally, since there are no histories in which $B$ arrives at time 2 and connects to its other neighbor $A$, we have directly $h_B(2) = 8$. This is easily verified with Table I.

The general formula for the total number of histories in which node $i$ arrived at time $t$ and connected to node $j$ is

$$g_{i \to j}(t) h_{j \to i} \binom{n-t-1}{n_{i \to j}-t}. \tag{7}$$

To compute $h_i(t)$, the total number of histories in which node $i$ arrived at time $t$ and connected to any of its neighbors, we simply sum this quantity over node $i$'s neighbors

$$h_i(t) = \sum_{j \in N_i} g_{i \to j}(t) h_{j \to i} \binom{n-t-1}{n_{i \to j}-t}. \tag{8}$$

All of this establishes that if we already know $g_{i \to j}(t)$ for each edge $i, j$ and time $t$ then we can calculate the quantities we want, $h_i(t)$. Of course we don't immediately know the $g_{i \to j}(t)$, and so for any of this to be of use we need a procedure to compute them. Fortunately, the logic is much the same.

Since $g_{i \to j}(t)$ is the number of histories in the branch containing $j$ in which $j$ arrives before time $t$, $g_{i \to j}(t+1) - g_{i \to j}(t)$ is the number in which $j$ arrives at exactly time $t$. This quantity is conceptually similar to $h_j(t)$ except instead of considering histories of the whole network, we only consider one branch. Restricting our attention to the branch containing $j$, once edge $i, j$ is removed, we can still apply Eq. (8), but we must be careful to remember that all the quantities are now defined relative to this branch. For example, while the original network had $n$ nodes, this branch will be a network with $n_{i \to j}$ nodes. We obtain

$$g_{i \to j}(t+1) - g_{i \to j}(t) = \sum_{k \in N_j \setminus i} g_{j \to k}(t) h_{i,k \to j} \binom{n_{i \to j}-t-1}{n_{j \to k}-t}, \tag{9}$$

where $h_{i,k\to j}$ is the number of histories in the branch containing $j$ once both edge $j,k$ and $i,j$ is removed. This quantity is almost, but not exactly, equal to $h_{k\to j}$. Whereas

$$h_{k\to j} = (n_{k\to j} - 1)! \prod_{l \in N_j \setminus k} \frac{h_{j\to l}}{n_{j\to l}!} \tag{10}$$

the quantity $h_{i,k\to j}$ is defined in the branch with $i$ already removed and thus

$$h_{i,k\to j} = (n_{k\to j} - n_{j\to i} - 1)! \prod_{l \in N_j \setminus \{i,k\}} \frac{h_{j\to l}}{n_{j\to l}!}. \tag{11}$$

Combining Eq. (10) with (11) and simplifying,

$$h_{i,k\to j} = \frac{h_{k\to j} n_{j\to i}! (n_{k\to j} - 1 - n_{j\to i})!}{h_{j\to i}(n_{k\to j} - 1)!}. \tag{12}$$

Equations (9), (10) and (12) together form a closed system, and can be solved with initial condition $g_{i\to j}(1) = h_{i\to j}$.

## COMPARISON WITH SEQUENTIAL MONTE-CARLO AND STRUCTURAL ESTIMATORS

In the main text, we introduce an efficient method to uniformly sample histories, $H$, that are consistent with an observed graph, $G$. We can estimate the expected value of any quantity $f$ using the Monte-Carlo method,

$$\langle f(H) \rangle = \sum_H f(H) P(H) \approx \sum_{s=1}^N f(H_s) P(H_s) \,,$$

where $P(H)$ is the probability of history $H$ under the considered growth model, and $\{H_s\}$ is a uniform sample of $N$ histories.

To illustrate its performance, we compared our approach with a state-of-the-art sequential Monte-Carlo (SMC) method [1, 2], combining sequential importance sampling (SIS) with a resampling procedure. Let us introduce the method briefly. An SIS scheme hinges on an easy to sample proposal distribution that factorizes as

$$Q(H) = \prod_{t=1}^{n-1} Q(H^t \mid H^{t-1}) \,,$$

where $H^t$ is a partial history of nodes up to time $t$. One can then estimate expected values,

$$\langle f(H) \rangle \approx \sum_{s=1}^N f(H_s) \frac{P(H_s)}{Q(H_s)} = \sum_{s=1}^N f(H_s) w(H_s) \,, \tag{13}$$

where histories $\{H_s\}$ are sampled from $Q(H_s)$, and $w(H_s)$ is the weight associated with history $H_s$.

However, this method alone often lead to poor estimates in practice. The root of the issue is that histories sampled from $Q(H_s)$ can be very unlikely and a few histories with much larger weights $w(H_s)$ can dominate the sum in Eq. (13), leading to a very small "effective" sample size. One standard way judge if this effective sample size is small is to consider whether

$$\text{ESS} = \frac{\left[\sum_{s=1}^N w(H_s)\right]^2}{\sum_{s=1}^N [w(H_s)]^2} \ll N \,.$$

A solution borrowed from the field of particle filtering is to perform the sampling of the histories in parallel, and at some point resample from the partial histories $\{H_s^t\}$ according to their weights $w(H_s^t)$. The new weights of the resampled histories is then reset to 1. This selection step creates duplicates for more likely histories while purging unlikely ones, which helps to keep a relatively high ESS.

The SMC method to which we compared our uniform sampling, consists of a snowball proposal distribution—proposing uniformly from the boundary—and resampling steps are performed whenever $\text{ESS}_t < N/2$, where $\text{ESS}_t$ is the effective sample size from the set of partial histories at time $t$ [1]. This is known as an adaptive SMC method.

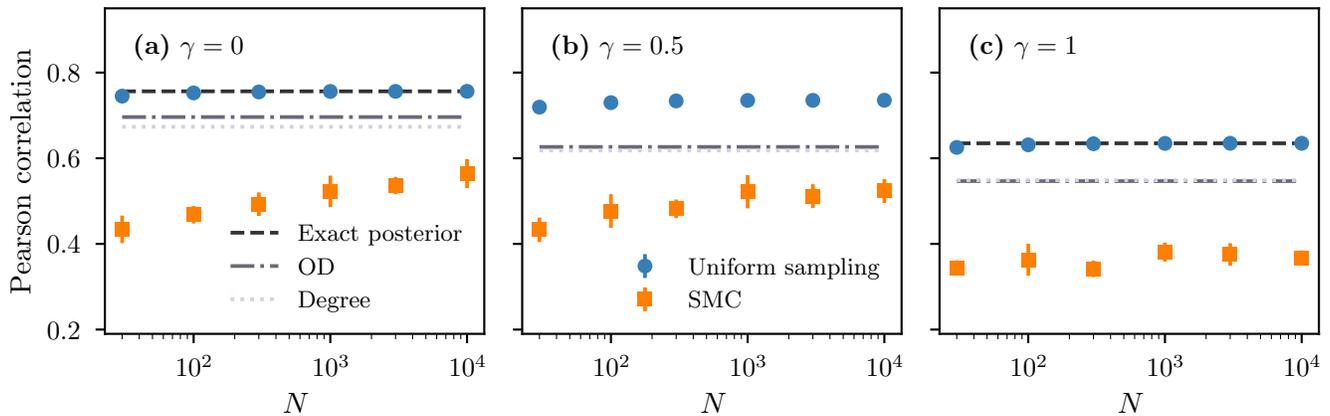

FIG. 2. Pearson correlation between the estimated arrival times of the nodes (from temporal reconstruction methods) and the ground truth. Each panel is associated with a single network of size $n = 500$ generated from a growth model with a non-linear degree-based attachment kernel $\propto k^\gamma$. All methods aim to recover the posterior means $\{\langle \tau_i(H) \rangle\}$ for all nodes $i$. The long dashed lines are the correlation of exact posterior, computed using Eq. (7) in the main text, with the ground-truth, while the short and intermittent dashed lines show the correlation attained by simpler, non-optimal, structural estimators [1]. The degree estimators rank node by decreasing degrees and the onion decomposition (OD) estimators rank nodes by centrality [3]. Markers are associated with estimates of the posterior means using sampling techniques : our uniform sampling method (circles) and sequential Monte-Carlo (squares). A set of estimates for $\{\langle \tau_i(H) \rangle\}$ is computed from $n$ sampled histories and a marker corresponds to the average correlation obtained over 10 realizations–error bars are the associated standard deviations.

In Fig. 2, we compare the ability of both methods to estimate the posterior means $\langle \tau_i \rangle$ for the arrival time of each node $i$, for a given number of samples $N$. We evaluate performance by the Pearson correlation coefficient of $\langle \tau_i \rangle$ with the real arrival times of the nodes. The estimator $\langle \tau_i \rangle$ has provably optimal performance for this task, although we compare to other estimators for reference [1].

We see in Fig. 2(a) and 2(c) that our uniform sampling method converges quickly to a history that is as correlated with the ground-truth as estimators that are calculated with the exact posterior mean, i.e., with Eq. (7) in the main text. In Fig. 2(b), the posterior $P(H)$ is no longer uniform, so we cannot compare with an analytic solution, but our method converges quickly nonetheless, and converges to a correlation that is superior to what naive methods find. In all cases, the SMC estimates improve only very slowly with increasing $N$, especially for the linear preferential attachment case [Fig. 2(c)], and remain inferior to direct uniform sampling. Other non-optimal structural estimators are shown as a comparison [1]: a temporal ranking of the nodes based on their degree and on the onion decomposition (OD) of the network [3].

## MODEL SELECTION

To select between two models one can uses Bayes factor,

$$K = \frac{P(G|M_1)}{P(G|M_2)}, \tag{14}$$

where $P(G|M)$ is the probability of generating network $G$ from model $M$ [4]. If $K \gg 1$ then $M_1$ is preferred, if $K \ll 1$ then $M_2$ is preferred, and if $K \approx 1$ then neither model is strongly favored.

Evaluating Bayes factor is generally non-trivial for growth models. The reason for this is that one must compute the model likelihood

$$P(G|M) = \sum_H P(G, H|M) \tag{15}$$

and the sum over histories may be prohibitively expensive. However, our Monte Carlo procedure uniformly samples possible histories and thus provides a method to evaluate this sum.

As an example, let's consider selecting between two models. Our first model is an underiected version of the redirection model of Ref. [5]. In this model, when each node arrives it uniformly selects an extant node. With

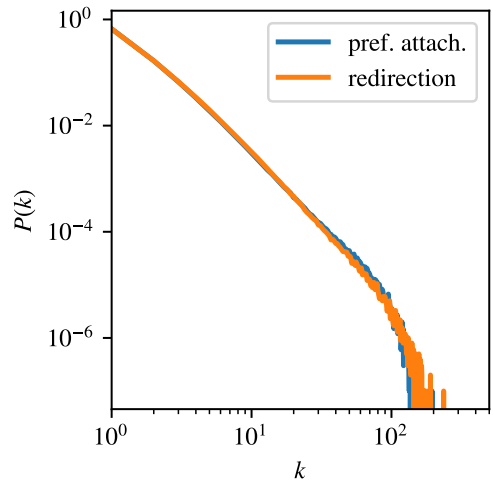

FIG. 3. Degree distributions for preferential attachment and redirection models. Ten thousand networks were sampled from both models, each with 2 048 nodes. For the redirection model, the redirection parameter was set to $r = 12/19$. Simulations show remarkably similar degree distributions for both models.

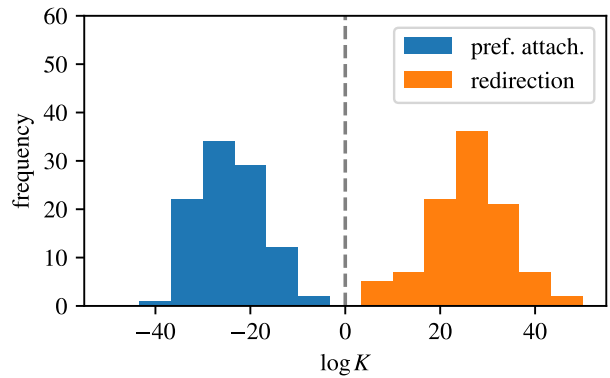

FIG. 4. We generated 100 networks with 2 048 nodes from both the preferential attachment and redirection models, and then computed Bayes factor for each one. Values of $K \ll 1$ are strong evidence in favour of preferential attachment, whereas $K \gg 1$ is strong evidence in favor of redirection. All networks were correctly identified—preferential attachment networks (blue) are to the left of the line, and redirection networks (orange) to the right.

probability $1 - r$ it then connects to that node, but with probability $r$ it instead connects to one of the neighbors of the chosen node. The second model we consider is the conventional preferential attachment model. Note, both of these models—the redirection model and preferential attachment—can have remarkably similar degree distributions. In Fig. 4 we show the result of simulating these models.

While both the redirection model and preferential attachment have similar degree distributions, they are not the same model and by computing Bayes factor we are able to tell them apart. We generated 100 networks with 2 048 nodes from each model and computed Bayes factor for each network. Without exception, Bayes factor correctly identified the true model. Thus, using these methods it is possible to distinguish the underlying mechanisms that generated a growing network.

* gcant@umich.edu
† guillaume.st-onge.4@ulaval.ca
‡ jean-gabriel.young@uvm.edu


\fi

\end{document}